# Replacing Noble Metals with Alternative Materials in Plasmonics and Metamaterials: how good an idea?

**Jacob B. Khurgin**

Johns Hopkins University, Department of Electrical and Computer Engineering
3400 N Charles St., Baltimore MD 21218 USA
jakek@jhu.edu

**Noble metals that currently dominate the fields of plasmonics and metamaterials suffer from large ohmic losses. New plasmonic materials, such as doped oxides and nitrides, have smaller material loss, and, using them in place of metals carries promise of reduced-loss plasmonic and metamaterial structures, with sharper resonances and higher field concentration. This promise is put to a rigorous analytical test in this work which reveals that having low material loss is not sufficient to have a reduced modal loss in plasmonic structures. To reduce the modal loss it is absolutely necessary for the plasma frequency to be significantly higher than the operational frequency. Using examples of nanoparticle plasmons and gap plasmons one comes to the conclusion that even in the mid-infrared spectrum metals continue to hold advantage over the alternative media. The new materials may still find application niche where the high absorption loss is beneficial, e.g. in medicine and thermal photovoltaics.**

## Introduction

Plasmonics and closely related to it field of metamaterials have been among the most active and exciting areas of optical sciences over the last decade [1]. The innate ability of plasmonic structures to squeeze optical energy into subwavelength volumes and thus achieve high degree of energy concentration allows one to enhance various linear, and especially nonlinear optical processes. This ability makes surface plasmon polariton (SPP) modes attractive candidates for use in applications like bio- and environmental sensing, photovoltaics, nonlinear optics, light sources, and all other areas where the extreme localization of electromagnetic fields can improve efficiency, speed, or, preferably both [1-3]. Propagating SPP's in sub-wavelength structures have been long promoted as the candidates for the future on-chip optical interconnects [4-6]. Metamaterials and metasurfaces relying on plasmonic resonances have been envisioned as the building elements of novel optical systems with widely extended functionality, such as super lenses [7]and optical cloaks and other transformational optics schemes [8].

Alas, the lofty promise of plasmonics and metamaterials has not been yet fulfilled and the chief culprit behind this disappointing turn of event is extremely high ohmic loss inherent to all metals, even the noble ones that are currently the mainstay of plasmonics and metamaterials. The salient feature of all the metals is the negative real part of the dielectric constant,

$$\varepsilon_r = \varepsilon_\infty \left[1 - \omega_p^2 / \left(\omega^2 + i\omega\gamma\right)\right] \qquad (1)$$

where $\varepsilon_\infty$ is the "background" dielectric constant of the bound electrons and ions, $\omega_p = (Ne^2/\varepsilon_0\varepsilon_\infty m)^{1/2}$ is the plasma frequency, $N$ is the carrier density, $m$ is effective mass, and $\gamma$ is the momentum scattering rate, responsible for the loss. The plasma frequency is typically in the ultraviolet region corresponding to wavelengths on the order of 150nm, while the momentum relaxation time is on the scale of $\gamma \approx 10^{14} s^{-1}$ For frequencies $\omega < \omega_p$ the real part of dielectric constant is negative and this fact allows subwavelength confinement in the SPP modes of various metal-dielectric structures.

The origin of sub-wavelength confinement in metal-containing structures can be understood from very simple energy conservation considerations [9]. In the dielectric cavity the energy oscillates between the electric and magnetic fields, just as in a resonant electric circuit the energy oscillates between the capacitor and inductor. One can also make a comparison with the mechanical oscillator by noting that electric field energy can be thought of as a potential energy, while the magnetic field energy plays the role of kinetic energy. Once the dimensions of the resonator become much smaller than wavelength, the magnetic field decreases and eventually becomes negligently small – well known quasi-static regime. The energy balance between the "potential" and "kinetic" energies can no longer be maintained and the mode ceases to exist as whatever the energy gets coupled into the cavity is radiated right back. But when the free electrons are present they possess the "true" kinetic energy, so now the energy can switch back and forth between the "potential" energy of the electric field and the kinetic energy of free carriers (plus some contribution from magnetic energy). The frequencies at which the energy balance can now be maintained are precisely the frequencies of the sub-wavelength SPP modes.

However, as explained in [9] once the energy is transferred into the form of kinetic motion carriers, it is unavoidably dissipated with the rate $\gamma$, hence the total energy of the SPP mode also decays at the rate commensurate with $\gamma \approx 10^{14} s^{-1}$. Thus higher confinement always leads to higher loss, at least in the visible and near IR ranges of optical spectrum.

Therefore in the last few years there has been a push to develop *alternative plasmonic materials*, such as strongly doped "conventional" semiconductors such as InGaAs [10] transparent oxides such as ITO, AlZnO [11] as well as nitrides (TiN, ZrN) and others [12-14]. One of the goals is to find a material with lower rate of the momentum relaxation, and indeed, this rate is lower at least in some semiconductors than in metals, a as shown in Table I, where for highly doped InGaAs relaxation rate is only $10^{13}$ s$^{-1}$ . The reason for it can be understood by realizing that with smaller density of electrons the Fermi layer in semiconductors lies lower than in metals, and there are fewer states into which the electrons can scatter. Unfortunately, the lower carrier density *N* also means lower plasma frequency and it causes that all the SPP resonances shift to longer wavelengths. Note that reduced effective mass in the semiconductor usually does not help much as the materials with small effective mass happen have the narrow bandgaps and correspondingly large background dielectric constant $\varepsilon_\infty$. In general, *γ increases and decreases in lock step with $\omega_p$.* since both are related to the density of states in the vicinity of the Fermi level.

So, reduced, in compassion to the noble metals, $\omega_p$ precludes using alternative plasmonic materials from use in visible and near IR spectral ranges, but at the same time, it appears to make them adequate, or perhaps even attractive when it comes to longer wavelengths. Indeed, if one considers a spherical "metallic" nanoparticle embedded into a dielectric with a dielectric constant $\varepsilon_d$ the resonant frequency of the lowest order localized SPP mode is $\omega_0 = \omega_p / (2 + \varepsilon_d)^{1/2}$ - hence in the noble metals it is usually in the blue or green regions of the spectrum, but with reduced plasma frequency the resonance can be easily shifted to let us say telecom range of 1300-1500nm or all-important mid-IR ranges of 3-5 and 8-12 micrometers. So, assuming that the momentum scattering in the alternative plasmonic materials is reduced, it makes them attractive for applications in mid-infrared plasmonics and metamaterials, as has been argued in [10-13].

Alas, at this time, experimental data do not support this optimistic projection as quality factors of the observed resonances in semiconductor nanostructures have been inferior to those seen in metallic designs. While part of the problem may lie in fabrication difficulties, recent numerical analysis [15] has shown that all said and done, when it comes to the loss, silver is still superior to all new plasmonic materials, including graphene. Yet no clear physical picture behind this "superiority" of Ag has emerged from that work. In this study we look into the same issue analytically and determine that the main reason for so-far-disappointing performance of "new" materials is simply *the low value of plasma frequency in them.*

**Kinetic and Magnetic Inductances**

To understand the loss one shall consider the energy balance in the plasmonic mode [9,16]. Half of the time, all the energy is stored in the form of the electrical energy whose density is $U_e = \frac{1}{2}\varepsilon_0\varepsilon_r E^2 \cos^2\omega t$. The other half time the energy is split between the magnetic energy with density $U_m = \frac{1}{2}\mu_0 H^2 \sin^2\omega t$ and kinetic energy of the collective motion free electrons $U_k = \frac{1}{2}Nmv^2 \sin^2\omega t$. The energy conservation, schematically shown in Fig.1 requires the total energy to be time independent and therefore $E_e = \int \bar{U}_e dV = E_m + E_k = \int(\bar{U}_m + \bar{U}_k)dV$, where the time-averaged value of electrical energy density is $\bar{U}_e = \frac{1}{4}\varepsilon_0\varepsilon_r E^2$ with similar expressions for the densities of kinetic and magnetic energies.

Neglecting the radiation (which can of course be useful as it provides means for coupling the energy in and out of the mode) neither electric nor magnetic energies are damped. But kinetic energy of moving carriers does get damped at twice the rate of the momentum damping, i.e. $2\gamma$. It is therefore not too difficult to see that the effective rate of energy loss in the mode is

$$\gamma_{eff} = \gamma \frac{E_k}{E_m + E_k} = \gamma \frac{\int \bar{U}_k dV}{\int (\bar{U}_m + \bar{U}_k)dV} \qquad (2)$$

Now, magnetic field in the mode is determined by the Maxwell equation $\nabla \times \boldsymbol{H} = i\omega\varepsilon_0\varepsilon_r \boldsymbol{E} + \boldsymbol{J}$. If one assumes that the fields are confined within characteristic dimension *a*, then in the absence of conductivity current the approximate relation between the magnitudes of two fields can be established as $H_1 \sim \varepsilon_0\varepsilon_r a\omega E = (an/\lambda)E/\eta_0$, where $n = \sqrt{\varepsilon_r}$ is the refractive index of dielectric, $\lambda$ is the wavelength in vacuum, and $\eta_0$ is the vacuum impedance. Therefore, as we have already noted, as the mode size decreases to less than $\lambda/n$ the magnetic field engendered by the time-dependent electric field (displacement current), $H_1$, gets progressively smaller eventually becoming negligibly small and so is its energy, $E_{m1} \sim (an/\lambda)^2 E_e$ is not sufficient to maintain the energy balance.

The field endangered by the conductivity current, $H_2 \sim I/a$ however does not explicitly depend on the relation between the mode size and wavelength. This field then can have sufficient magnetic energy $E_{m2} \sim \frac{1}{4}\mu_0 \int \bar{H}_2^2 dV = \frac{1}{4} L_m I^2$ to reduce the need for the kinetic energy $E_k = \int \bar{U}_k dV = \frac{1}{4} L_k I^2$, where we have introduced $L_m$ as a "conventional" magnetic inductance and $L_k$, the *kinetic inductance*. The kinetic induction is a measure of the inertia of the free electron gas which causes the current to lag in phase behind the voltage.

With these definitions (2) the main expression for the effective loss becomes

$$\gamma_{eff} = \gamma(1 + L_m/L_k)^{-1} \qquad (3)$$

This effective loss or, rather the quality factor $Q = \omega/\gamma_{eff}$, determines the maximum field enhancement attainable in the plasmonic structure; hence it is based on the relation (3) that we shall now perform comparison of different plasmonic materials

**Loss Scaling in Localized SPP's**

Consider now the SPP mode around elliptical nanoparticle which we can approximate as a cylinder of length *l* and diameter *a* (Fig2.a). The length *l* can be adjusted to achieve an SPP resonance at a given wavelength $\lambda$ no matter what is $\omega_p$. The electric field, and hence current, penetrates the metal by the skin depth, roughly $\lambda_p/2\pi$ where $\lambda_p = 2\pi c/\omega_p$ is the plasma wavelength. Therefore the effective cross-section in which the current is contained can be approximated as

$$S_{eff} = \frac{\pi a^2}{4}\left[1 + \left(\frac{\pi a}{2\lambda_p}\right)^2\right]^{-1/2} \qquad (4)$$

Clearly, if the diameter of the nanoparticle is much smaller than the skin depth, then the field permeates the nanoparticle and $S_{eff} = \pi a^2/4$, while in large diameter nanoparticles the current is contained in the cylindrical region of skin depth thickness with area $S_{eff} = a\lambda_p/2$.

The total kinetic energy is $E_k = \int U_k dV = \frac{1}{2} Nmv^2 S_{eff} l = \frac{1}{2} mI^2 l/Ne^2 S_{eff}$, while the current is $I = \int eNvdS = eNvS_{eff} l$ and using definition of plasma frequency we obtain

$$E_k = \tfrac{1}{2} mI^2 l/Ne^2 S_{eff} = \tfrac{1}{2}\varepsilon_0 I^2 l/\omega_p^2 S_{eff} \qquad (5)$$

From this expression the for the kinetic inductance easily follows.

$$L_k = \mu_0 l \lambda_p^2 / 4\pi^2 S_{eff} \tag{6}$$

As one can see, the actual wavelength of light is absent from (6) and the kinetic inductance depends only on the ratio between the plasma wavelength and the mode dimension $a$. One can also find the resistance of the nanoparticle as

$$R = \gamma \frac{ml}{e^2 N S_{eff}} = \gamma L_k, \tag{7}$$

which allows us to re-write (3) as simply $\gamma_{eff} = R/L$, where $L \equiv L_k + L_m$ which is exactly what one would obtain from simple resonant LC circuit consideration

While kinetic inductance exhibits strong dependence on the transverse dimensions, the magnetic inductance show only very weak dependence on the size $a$,

$$L_m = \frac{\mu_0}{2\pi} l \left[ \log\left(\frac{4l}{a}\right) - 1 \right] \tag{8}$$

The expression in the square brackets changes from 1 and 2 as the ratio $(l/a)$ increases from 1.5 to 5 and thus can be neglected in our order-of-magnitude analysis. Thus we obtain a very simple relation between the two inductances

$$\frac{L_m}{L_k} \approx \frac{2\pi S_{eff}}{\lambda_p^2} = \frac{\pi a^2}{2\lambda_p^2}\left[1 + \left(\frac{\pi a}{2\lambda_p}\right)^2\right]^{-1/2} \tag{9}$$

and then the value of effective loss rate can be found as

$$\gamma_{eff} \approx \gamma / \left(1 + 2x^2/\sqrt{x^2+1}\right) \tag{10}$$

where $x = \pi a / 2\lambda_p$.

The effective loss, and therefore broadening and maximum field enhancement do not depend on the wavelength of light, but only on the plasma wavelength. It is conceivable therefore that metal with larger γ and smaller λ_p would have smaller loss than a doped semiconductor. As one can see from the Fig.2 , where the effective loss is plotted for the case of subwavelength structure $a=\lambda/10$. To illustrate the result (10) we consider an elliptical particle with a dimeter $a = \lambda/10$ made from either gold or InGaAs doped to the degree of $2 \times 10^{19} cm^{-3}$. As follows from the Table I, the scattering rate in the semiconductor is five time smaller than in gold, but the plasma frequency is smaller by a factor of 10. As mentioned above, the resonant frequency of the SPP can be tuned by changing the aspect ratio of the nanoparticle, hence the length $l$ of the Au nanoparticle is always longer than that of semiconductor one. The results are shown in Fig.3. For gold nanoparticle the SPP resonance can be tuned anywhere in the visible and IR range. In the visible range the loss in of the Au SPP is just as high as the scattering rate in gold ($\gamma_{eff} \sim 10^{-14} s^{-1}$), but in the mid-IR range the diameter $a = \lambda/10$ exceeds $\lambda_p$ and the lass rate decreases to $\gamma_{eff} < 10^{-13} s^{-1}$ beyond $\lambda = 3\mu m$ i.e. less than the scattering rate in the semiconductor. For the InGaAs nanoparticle the SPP mode exits only at $\lambda \geq 3\mu m$, and, since for InGaAs in this range $a = \lambda/10 > \lambda_p$ the loss of the semiconductor SPP mode always remains higher than in the Au SPP mode, despite the fact that scattering rat ein the semiconductor is lower. Note, that in the end at long wavelengths with $a >> \lambda_p$ the effective loss becomes simply $\gamma_{eff} \approx \gamma \lambda_p / \pi a$, hence the *proper figure of merit for the material should be $\omega_p/\gamma$ and not often used $\omega/\gamma$. According to Table I the proper figure of merit is always higher for metals than for any of semiconducting materials.*

One can also consider a split ring resonator [17](Fig.2b) with the circumference $l$ and the cross-section of the wire $\pi a^2 / 4$. The resonance frequency in it can always be adjusted by properly choosing the size of the gap $d$. The magnetic inductance of the loop once again does not depend on the cross-section, $L_m = \mu_0 l / 4$ and the result almost identical to (10), $\gamma_{eff} \approx \gamma / \left(1 + \pi x^2 / \sqrt{x^2 + 1}\right)$ can be obtained, indicating the generality of our approach.

Note that one of the main arguments behind looking for alternative materials is that in order to achieve high degree of confinement in the plasmonic structure the permittivity of metal and dielectric should have roughly equal magnitudes and opposite signs, $\varepsilon_m \approx -\varepsilon_d$, and therefore to achieve good confinement at long wavelength the plasma frequency should be "red shifted" there [10-13]. This argument, however, is totally flawed – no matter how large is $|\varepsilon_m|$ one can always achieve high degree of confinement by shifting the resonant SPP frequency via increased capacitance by designing the structure with a sufficiently small gap, changing eccentricity of the elliptical nanoparticle or using nanoshells and dimers. If anything, reducing the capacitance and shifting the SPP resonance to the blue has always presented the problem, but not shifting it to the red and beyond!

**Loss Scaling in Gap SPP waveguides.**

One can extend this comparison to the case of the gap SPP [18-21] confined within the dielectric of thickness $a$ placed between the two layers of metal or other plasmonic material as shown in Fig.1 c. Unlike simple SPP on the metal/dielectric interface, the gap SPP allows one to achieve sub-wavelength confinement for very broad range of frequencies. The tangential magnetic and electric fields in SPP can be written as

$$H_y = \begin{cases} H_0 \cosh(q_d x) e^{j\beta z} & x < a/2 \\ H_0 \cosh(q_d a/2) e^{-q_m(x-d/2)} e^{j\beta z} & x > a/2 \end{cases}, \quad (11)$$

and

$$E_z = \begin{cases} -j(q_d/\omega\varepsilon_d) H_0 \sinh(q_d x) e^{j\beta z} & x < a/2 \\ j(q_d/\omega\varepsilon_m) H_0 \cosh(q_d a/2) e^{-q_m(x-d/2)} e^{j\beta z} & x > a/2 \end{cases} \quad (12)$$

respectively. The propagation constant $\beta$ and the decay coefficients $q_d$ and $q_m$ are related as

$$\begin{aligned} \varepsilon_m k_0^2 &= \beta^2 - q_m^2 \\ \varepsilon_d k_0^2 &= \beta^2 - q_d^2 \end{aligned} \quad (13)$$

and $k_0 = \omega/c$, and from (13) one obtains

$$q_m^2 = q_d^2 + (\varepsilon_d - \varepsilon_m) k_0^2 \quad (14)$$

Applying the boundary condition for the continuity of the tangential component of electric field immediately yields

$$\frac{q_m}{\varepsilon_m} \cosh(q_d a/2) = -\frac{q_d}{\varepsilon_d} \sinh(q_d a/2) \quad (15)$$

From (14) and (15) the recursive relation for the decay constant in dielectric

$$q_d = -\frac{\varepsilon_d}{\varepsilon_m} \frac{\sqrt{q_d^2 + (\varepsilon_d - \varepsilon_m) k_0^2}}{\tanh(q_d d/2)} \quad (16)$$

For the gap case of a thin subwavelength gap with $\beta \sim k_0 \varepsilon_d^{1/2}$ we have $q_d d/2 \ll 1$ and for operational wavelength far from the surface plasmon resonance $|\varepsilon_m| \gg \varepsilon_d$ and one obtains

$$q_d \approx k_0 \frac{\varepsilon_d}{\sqrt{-\varepsilon_m}} \frac{1}{q_d d/2} \quad (17)$$

and using Drude formula for the dielectric $\varepsilon_m = 1 - \omega_p^2/(\omega^2 + i\gamma) \approx -\omega_p^2/(\omega^2 + i\gamma)$ constant of metal, the decay coefficient in the dielectric medium is

$$q_d \approx \left(\frac{2k_0\omega(1+i\gamma/2\omega)}{a\omega_p}\right)^{1/2} \varepsilon_d^{1/2} = \left(\frac{2\lambda_p(1+i\gamma/2\omega)}{\pi a}\right)^{1/2} k_0 \varepsilon_d^{1/2} \qquad (18)$$

Substituting this into (13) we finally get the dispersion relationship

$$\beta \approx k_0 \varepsilon_d^{1/2}\left(1+\lambda_p/\pi a\right)+i\frac{\gamma\lambda_p\varepsilon_d^{1/2}}{2\pi ac} \qquad (19)$$

After some algebra one can find the dispersive relation for the plasmon under assumption that effective propagation constant β is not significantly different from the wavevector in dielectric $k_d$ which is a typical practical arrangement.

$$\beta = k_d + \frac{\omega}{\omega_p a}(1+j\gamma/2\omega) \qquad (20)$$

The propagation length then can be found as

$$L_{prop} = \frac{1}{2\beta} = \frac{\pi ca}{\lambda_p \gamma n} \qquad (21)$$

where n is the index of dielectric. Once again, the loss depends only on the ratio $\pi a/\lambda_p$ and the figure of merit (FOM) should be the same expression $\omega_p/\gamma$ obtained above for the localized SPP.

Using the Drude expression of dielectric constant once again one obtains

$$FOM = \frac{\omega_p}{\gamma} \approx \frac{|\varepsilon_m'|^{3/2}}{\varepsilon_m''}, \qquad (22)$$

where $\varepsilon_m'$ and $\varepsilon_m''$ are the real and imaginary parts of dielectric constant of the metal. *It is easy to see(Table I) that the noble metals have higher FOM than most if not all "new" plasmonic material all the way to far IR range*, just as numerical simulations in [15] have shown.

To illustrate this fact we perform full modeling of the dispersion and propagation length in SPP plasmons for gold and InGaAs waveguides with SiO$_2$ core. The results are shown in Fig. 5. The propagation constant (effective) index $\beta$ of the gap SPP shown in Fig. 5 a and c stays equal to $k_d$ for as long as $a > \lambda_p/2\pi$ and then gradually increases as the field starts penetrating the metal (semiconductor). The propagation length indeed increases proportionally to the gap thickness and for the metal (Fig.5b) it is almost independent on the wavelength in full agreement with (21) while for InGaAs (Fig.5d) the propagation length does depend on the wavelength. This can be explained by the fact that when wavelength approaches the plasma wavelength the field penetrates deep inside the semiconductor. At any rate, comparing Fig.5b and 5d one can see that propagation length in metal always exceeds that in semiconductor despite higher material loss in the metal.

## Conclusions

We have compared performance of the alternative plasmonic materials in the IR region with that of the metals using a simple fully analytical model. We have shown that the proper figure of merit, defining loss, broadening and field enhancement should be the ratio of plasma frequency $\omega_p$ and material loss $\gamma$. *In laymen terms it indicates that it is always preferable to have many electrons moving relatively slow rather than a relatively few electrons moving very fast.* The figure of merit of all the alternative materials is worse than that of Ag and even Au. Therefore one cannot expect better performance from the new materials in the "traditional" plasmonic applications, where the loss, propagation length, or degree of field enhancement is not the most important factor. It is also important to realize that there exist important application niches for the alternative plasmonic materials due to them having a higher melting point [22], their compatibility with existing (Si or III-V) technologies, or simply based on the price and availability considerations. Furthermore, some applications in medicine [23], photocatalysis [24], and thermal photovoltaics [25] actually require high loss and broad resonances and it is in these applications that the alternative plasmonic materials may find their practical use.

## Acknowledgement
The author wishes to acknowledge generous of US Army Research Office (grant W911NF-15-1-0629 ) and stimulating discussions with his colleague Prof. P. Noir of Johns Hopkins University

**References**

1. S. A. Maier, *Plasmonics: Fundamentals and Applications*, Springer, Boston, MA **2007**.
2. M. I. Stockman, "Nanoplasmonics: past, present, and glimpse into future," Opt. Express **19**(22), 22029–22106 (2011)
3. M. Pelton, "Modified spontaneous emission in nanophotonic structures", Nat. Photon. **9**, 427-435 (2015).
4. E. Ozbay, "Plasmonics: merging photonics and electronics at nanoscale dimensions," Science **311**, 189 (2006).
5. Z. Han and S. I. Bozhevolnyi, "Radiation guiding with surface plasmon polaritons", Rep. Prog. Phys. **76**, 016402 (2013).
6. J.A. Conway, S. Sahni, and T. Szkopek, "Plasmonic interconnects versus conventional interconnects: a comparison of latency, crosstalk and energy costs," Opt. Express **15**, 4474-4484 (2007)
7. X. Zhang and Z. Liu Superlenses to overcome the diffraction limit", Nature Materials **7**, 435 - 441 (2008)
8. C. Atre, A. García-Etxarri, H. Alaeian, J. A. Dionne, "A Broadband Negative Index Metamaterial at Optical Frequencies", Advanced Optical Materials, **1**, 327-333 (2013)
9. J. B. Khurgin, "How to deal with the loss in plasmonics and metamaterials", *Nature Nanotechnology* **10** 2-6, (2015)
10. S. Law, D. C. Adams, A. M. Taylor, and D. Wasserman, "Mid-infrared designer metals," Opt. Express **20**, 12155-12165 (2012)
11. A. Boltasseva, H. A. Atwater, "Low-Loss Plasmonic Metamaterials", *Science* **2011**, *331*, 290
12. A. Boltasseva, "Empowering plasmonics and metamaterials technology with new material platforms", *MRS Bulletin,* 39, **461** (2014)
13. G.V. Naik, V. M. Shalaev, and A. Boltasseva," Alternative Plasmonic Materials: Beyond Gold and Silver", Adv. Mater, **25**, 3264 (2013)
14. S. A. Gregory, Y. Wang, C.H. de Groot, and Otto L. Muskens," Extreme Subwavelength Metal Oxide Direct and Complementary Metamaterials" ACS Photonics **2**, 606-614 (2015)

15. B. Dastmalchi, P. Tassin, T. Koschny, C. M. Soukoulis, "A New Perspective on Plasmonics: Confinement and Propagation Length of Surface Plasmons for Different Materials and Geometries", Advanced Optical Materials, **4** pp 171-184 (2016)
16. J. B. Khurgin, G. Sun, "Scaling of losses with size and wavelength in nanoplasmonics" Appl. Phys. Lett, **99**, 211106 (2011)
17. J. Zhou, T. Koschny, and C. M. Soukoulis, "Magnetic and electric excitations in split ring resonators," Opt. Express **15**, 17881-17890 (2007)
18. K. Tanaka, M. Tanaka, and T. Sugiyama, "Simulation of practical nanometric optical circuits based on surface plasmon polariton gap waveguides," Opt. Express **13**, 256-266 (2005).
19. S. H. Chang, T. C. Chiu, and C.-Y. Tai, "Propagation characteristics of the supermode based on two coupled semi-infinite rib plasmonic waveguides," Opt. Express **15**, 1755-1761 (2007). 7.
20. I. V. Novikov and A. A. Maradudin, "Channel polaritons," Phys. Rev. B **66**, 035403 (2002).
21. S. I. Bozhevolnyi, V. S. Volkov, E. Devaux, and T. W. Ebbesen, "Channel plasmon-polariton guiding by subwavelength metal grooves," Phys. Rev. Lett. 95, 046802 (2005).
22. W. A. Challener, Chubing Peng, A. V. Itagi, D. Karns, Wei Peng, et al, "Heat-assisted magnetic recording by a near-field transducer with efficient optical energy transfer", Nature Photonics **3**, 220 - 224 (2009)
23. L. R. Hirsch, R. J. Stafford, J. A. Bankson, S. R. Sershen, B. Rivera, R. E. Price, J. D. Hazle, N. J. Halas, and J. L. West. Nanoshell-Mediated Near-Infrared Thermal Therapy of Tumors under Magnetic Resonance Guidance. P. Natl. Acad. Sci. USA **23**, 13549-13554 (2003)
24. X. Zhang, Y. Lim Chen, R. S. Liu and D. Ping Tsai," Plasmonic Photocatalysis", Rep. Prog. Phys., **76**, 046401 (2013)
25. Y. Guo, S. Molesky, H.Hu, C. L. Cortes, Z. Jacob Z" Thermal excitation of plasmons for near-field thermophotovoltaics". Appl. Phys. Lett. **105**, 073903 (2014)

**Table 1**

**Parameters of plasmonic materials – metals and doped semiconductors (From Refs 9-14)**

| Material | $\omega_p$(eV) | $\gamma$(eV) | $\omega_p/\gamma$ |
|---|---|---|---|
| Ag | 9.2 | 0.02 | 460 |
| Au | 8.9 | 0.07 | 127 |
| Al | 12.7 | 0.13 | 98 |
| AZO(2%) | 1.74 | 0.045 | 40 |
| ITO(10%) | 1.78 | 0.15 | 12 |
| TiN | 8 | 0.188 | 45 |
| InGaAs(doping $2\times10^{19}$ cm$^{-3}$) | 0.8 | 0.014 | 57 |

**Figure Captions**

Fig.1 Energy Balance in the plasmonic mode. Every half-period the energy is transferred from the electrical energy $E_e$ into the sum of magnetic energy $E_m$ and kinetic energy of free carriers $E_k$. Magnetic energy has two components – $E_{m1}$ engendered by the displacement current and $E_{m2}$ engendered by the conductivity current **J**.

Fig.2 (a) Elliptical plasmonic nanoparticle and the sketch of the electric field penetrating it. (b) Split ring resonator.

Fig.3 Effective loss in the elliptical nanoparticle made from silver and semiconductor (ITO) vs. the wavelength

Fig.4 Gap SPP and electric field in it

Fig.5 Propagation constant (a,b) and propagation length (c,d) at different wavelengths as a function of gap thickness for the metal and semiconductor waveguides respectively.

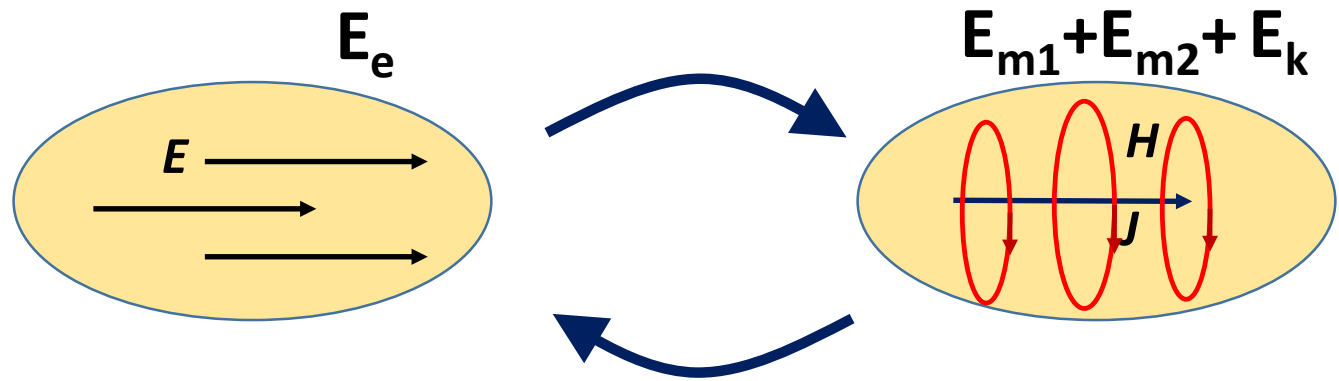

**Fig 1**

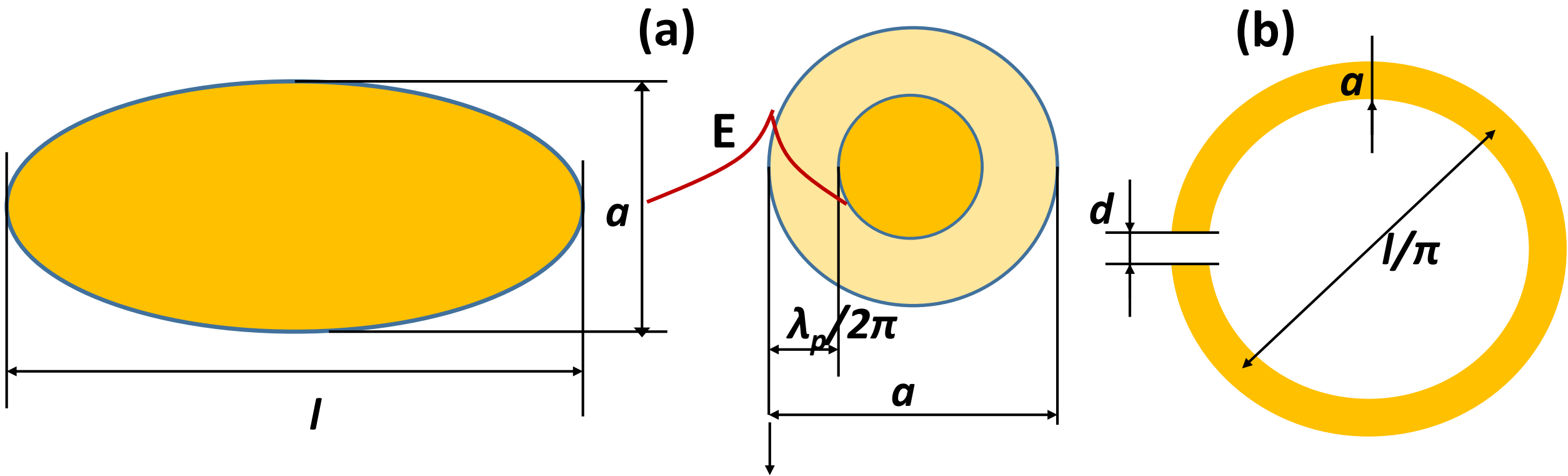

Fig 2

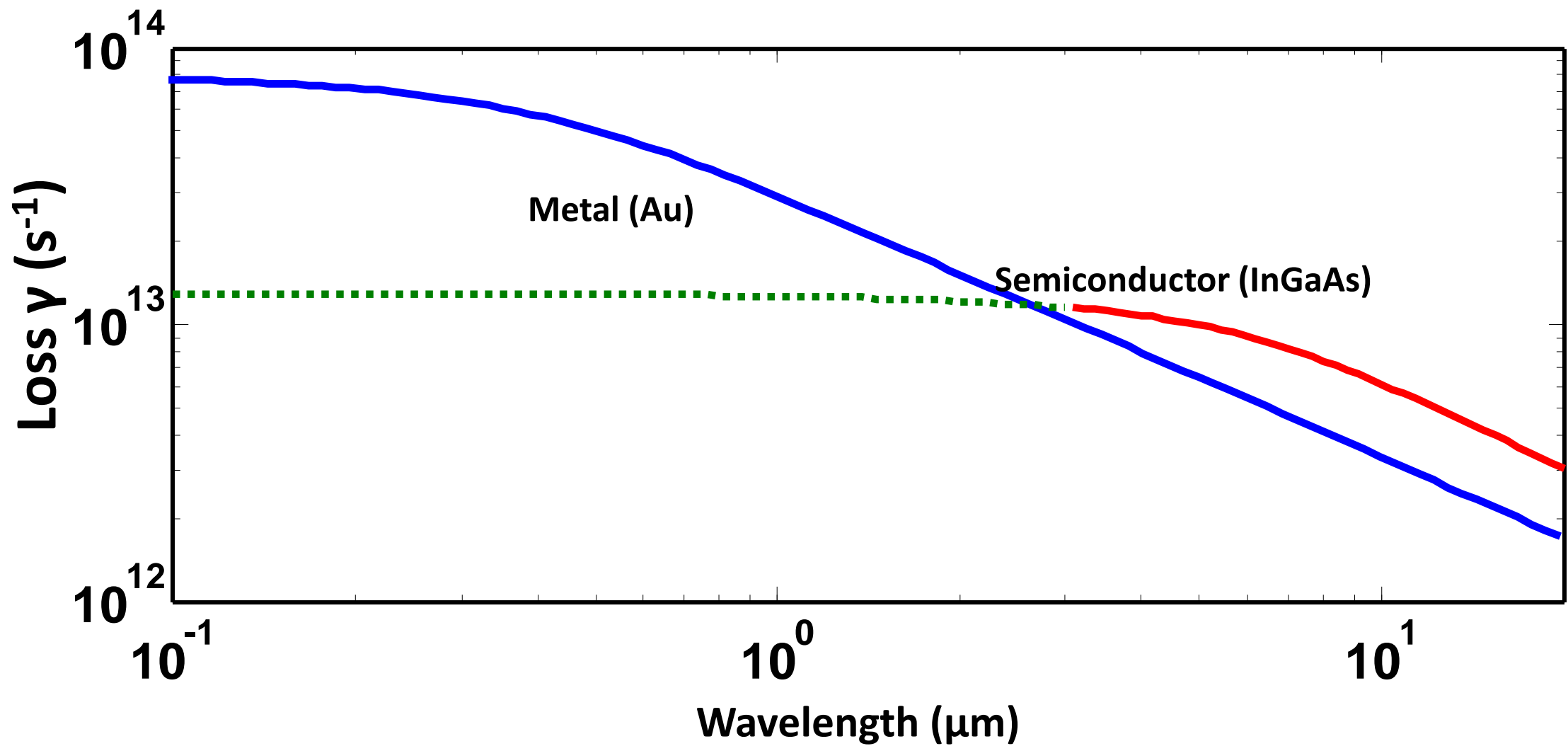

Fig.3

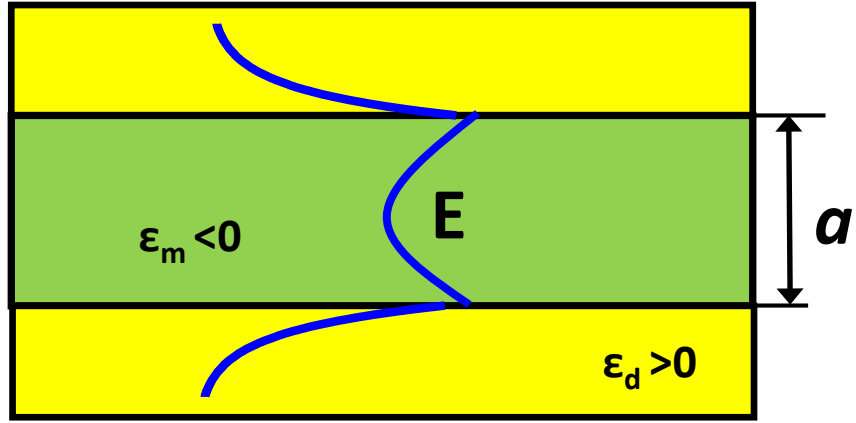

Fig.4

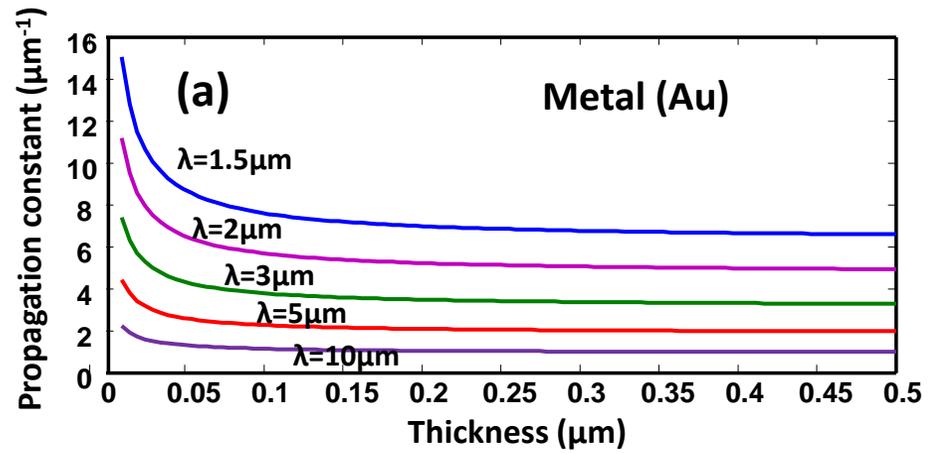
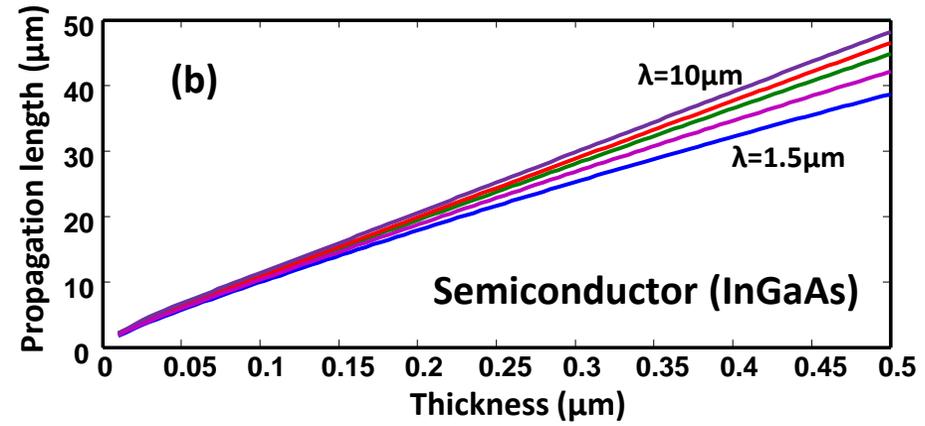
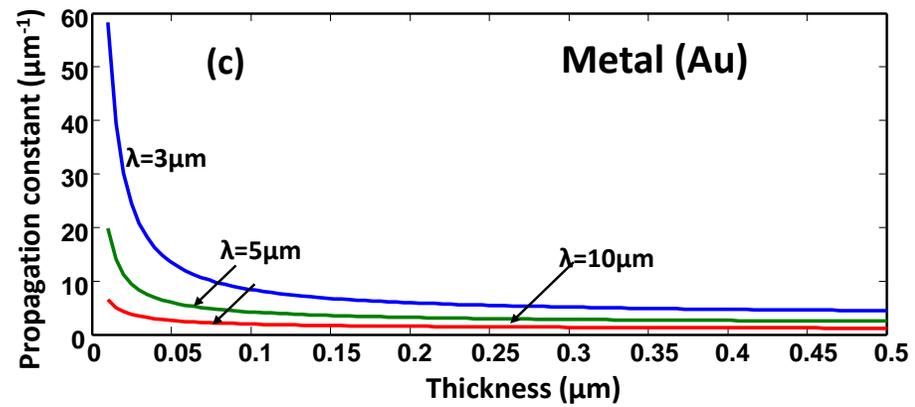
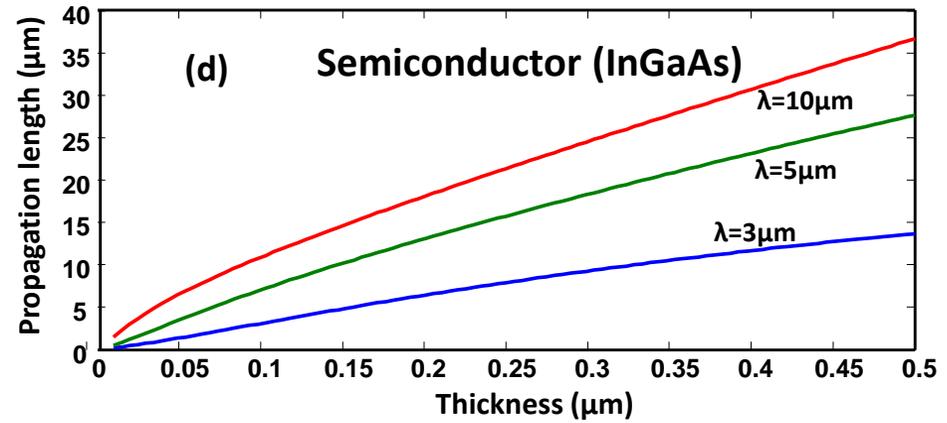

Fig.5